\documentclass[showkeywords,preprint,pre,onecolumn]{revtex4}
\usepackage{graphicx}
\usepackage{amsfonts}
\usepackage{amssymb}
\usepackage{latexsym}
\usepackage{ifpdf}
\usepackage{natbib}

\pdfoutput=1
\begin{document}

\title{Towards a model of the human society: A theoretical solution of the cooperation problem}

\begin{abstract}
 
\bigskip 

In this paper we address the cooperation problem in structured populations by considering the prisoner's dilemma game as metaphor of the social interactions between individuals with imitation capacity. We present a new strategy update rule called \emph{democratic weighted update} where the individuals behavior is socially influenced by each one of their neighbors. In particular, the capacity of an individual to socially influence other ones is proportional to its wisdom which is defined by its successful in the game. When in a neighborhood there are cooperators and defectors, the focal player is contradictorily influenced by them and, therefore, the effective social influence is given by the difference of the total wisdom of each strategy in its neighborhood. First, by considering the growing process of the network and neglecting mutations we show the evolution of highly cooperative systems. Then, we broadly shown that the social influence allows to overcome the emergence of mutants into highly cooperative systems. In this way, we are able to conclude that considering the growing process of the system, individuals with imitation capacity and the social influence the cooperation evolves. Therefore, here we present a theoretical solution of the cooperation problem among unrelated individuals with imitiation capacity.

\bigskip 


\end{abstract}

\author{Ignacio Gomez Portillo \footnote{E-mail: ignaciogportillo@gmail.com}}
\affiliation{Grup de F\'isica Estad\'istica, Departament de F\'isica, Universitat Aut\'onoma de Barcelona, 08193 Barcelona, Spain.}
\maketitle

\section{Introduction}

Cooperation is ubiquitous in every form of life \cite{G1,G2,G3,G4}, ranging from genes to societies such as the human one. The emergence of cooperation is an evolutionary transition that increases the complexity of life by forming new biological organisms from other ones. It is interesting to note that some biological individuals such as the human beings and the ants have a dual nature, they are part of a more complex organism at the same time that an organism more complex than its parts. This self-similar feature shows that cooperation is a fractal property of life. Besides, it is important to note that the cooperation among organisms is temporary since every form of life dies. In this sense, the cooperative organisms must to develop reproductive capacity in order to evolve beyond its life. Therefore, understanding the mechanisms that allow the formation, persistence and reproduction of cooperative systems is essential. The evolution of cooperation has been widely considered  \cite{G5,G6,G7,G8,G9,G10} in literature. In this long and successful tradition, lot of attention has been paid to better understand the evolution of our society. Although several proposed models \cite{G11} allow explaining the cooperation of many actual situations, we still cannot solve the problem in a general way. Therefore, we have to keep looking for answers to develop a theoretical model of our society.

\bigskip

Evolutionary game theory \cite{G12,G13} has proven to be an appropriate theoretical framework to formally address the cooperation problem. In this theory, the most important prototypical social interactions \cite{G14} are represented by a game where each individual adopts a strategy to play against its opponent. In order to consider Darwinian theory, the system evolves favoring the replication of the successful strategies. In particular, the Prisoner's Dilemma has been the most widely studied game as metaphor of the cooperation problem. In this, each player adopts one of the two possible strategies, cooperation or defection. When in an interaction both individuals cooperate each receive a payoff $R$ and other one $P$ under mutual defection. If one cooperates and the other defects, the former receives $S$ and the second $T$. The game is completely defined when $T>R>P>S$. Under these conditions and in a one-shot game, it is better to defect no matter the strategy adopted by the opponents. Thus, in fully connected systems defection always has the highest reproduction rate. Therefore, the system evolves decreasing the fraction of cooperators to extinction. However, individuals of actual cooperative systems have local information of the system instead global one. They are placed in a structured population interacting with just some other ones. Taking this into account, several studies have been performed  \cite{R1,R2,N1,N2,N3,N4,N5,N6,N7,N8,N9,N13,N10,N11,N12} considering different properties of actual networks interactions. In this way, it has been shown that some features such as locality \cite{N1} and degree heterogeneity \cite{N3,N4,N5} could be \cite{N13,N10,N11} of great importance for the evolution of cooperation. However, given the high benefit-cost ratio (see next section) required for cooperation to evolve respect to the one observed in nature, which is particularly evident when the system has a high average number of connections \cite{N12}, the problem is not completely solved just considering the structure of actual systems. In order to overcome this limitation, some features such as individuals who can distinguish their opponents \cite{W1}, rewiring process introduced through coevolutionary dynamics \cite{C1} and multiplex networks \cite{MN} have been considered showing great importance for the evolution of cooperation.  

\bigskip

Recently, it has proposed \cite{I1,I2} a new formulation of the cooperation problem taking into account two general features of actual cooperative systems, \emph{i.e.} the growing process and the possibility of mutations. By considering individuals with imitation capacity and neglecting mutations, they have shown under very low conditions mechanisms to build highly cooperative systems of any size and topology. In particular, the minimal benefit-cost ratio $r$ required for cooperation to evolve approaches to the theoretical minimum $r=1$ when the average number of links of the system increases. Remarkably, the growing process generates locality and degree heterogeneity that, as we previously stated, seem to be important features for the evolution of cooperation. However, from this process other features of actual systems emerge drastically improving the required conditions for cooperation to evolve. On the one hand, the system have a high level of cooperation in all stages instead half or less of the population normally considered as initial condition in already formed systems \cite{R1,R2}. On the other hand, defectors inhabit the less linked parts of the system which is the optimal arrangement for the cooperation \cite{CI}. In this way, they have shown that the growing process is essential for the evolution of cooperation when the individuals have imitation capacity as the human being. Nevertheless, the emerging cooperative systems require a very high benefit-cost ratio \cite{CI} to overcome the emergence of mutant defectors in highly linked individuals of the system. In this way, the new organism is alive just until highly connected mutants defectors appear into the system. In this sense, the longevity and the system size strongly dependent of the mutation rate of individuals. However, there are organisms such as the human society whose size and longevity cannot be explained completely by the model. Therefore, in order to show theoretically the very existence of this kind of systems, it is required to introduce new features to the model allowing to the system overcomes the apparition of mutants defectors. In this paper, we address this problem performing considerations over the strategy update rule by which individuals adopt their strategy.

\bigskip

Through the evolution of the system, the individuals adopt a strategy by evaluating the information provided by the environment or by mutation. When the problem is considered over structured populations, each individual have information from its neighborhood, \emph{i.e.} itself and its neighbors. Normally, players are considered without memory or prediction capacity. Therefore, the information of each individual is restricted to payoffs and strategies from the last round of its neighborhood. Generally, the strategy updates of each individual are performed by comparison of the payoff of the focal player with the one of a neighbor, who can be the most successful or a randomly selected one. In this way, the individuals with more than one link choose their strategy neglecting part of the information available. However, a strategy update where the whole neighborhood affects simultaneously the behavior of the focal player seems to be more realistic. Taking this as motivation, a new strategy update rule \cite{S1} has been recently proposed where the focal player evaluates its strategy by comparing the average payoff of each strategy in the neighborhood. Studying the new rule over already formed regular lattices, they have shown a significant increment in the survivability of cooperators with respect to the one obtained by a pairwise comparison rule. Although the average payoff of each strategy allows taking into account all the payoffs simultaneously, this strategy update rule is still neglecting part of the information available since the abundance of each strategy in the neighborhood is not taken into account. However, the last is very important in order to exploit properly the information available from individuals. Fortunately, this has been recently taken into account \cite{S2} through the learning activity of players by considering that it increases with the number of neighbors with different strategy. In particular, they have defined the wisdom of each strategy in a neighborhood as its abundance. In this way, they have shown that the neighborhood wisdom enhances the survivability of cooperation over already formed regular lattices. This occurs by a dynamically decelerated invasion process, by means of which interfaces separating different domains remain smooth and defectors therefore become unable to efficiently invade cooperators.

\bigskip

Here, we present a new strategy update rule, that we call \emph{democratic weighted update}, where the average payoff and the abundance of each strategy in the neighborhood of the focal players is taken into account simultaneously. It is important to state that we define the wisdom of individuals in a very different way than in \cite{S2}. We consider the wisdom of each individual as a combination of its strategy and payoff instead just its strategy. In particular, we define the wisdom of each individual as proportional to its payoff. In this way, individuals with high payoff are wiser than those with low one. Thus, the wisdom of the system is heterogeneously distributed among individuals instead of being homogeneous. We justify this consideration by the following. First, from the perspective of a focal player, the neighbors with high payoff seems to be better players than those with low ones and, therefore, they are more reliable. In this sense, the payoff of each individual can be interpreted as a measure of its reputation regardless of its strategy. Second, from the statistical point of view, when the connectivity of the system is heterogeneous the payoff of an individual normally increases with its degree. Thus, individuals with high payoff have, in average, more information of the system than individuals with low one and, therefore, their information is more reliable. Otherwise, we consider that the behavior of individuals is socially influenced by the wisdom of every individuals of their neighborhood. When the total wisdom of the neighborhood is divided by the existence of both strategies, the focal player is influenced by these contradictory knowledge to opposite directions. Hence, the effective social influence over the focal player is defined by the difference between the total wisdom of each strategy instead by the absolute one considered through the learning activity. Through the work we broadly shown that the social influence allows to structured highly cooperative systems overcome the emergence of mutant defectors. In this way and considering \cite{I1,I2}, we conclude that in structured growing system whose individuals have imitation capacity and their behavior is affected by social influence cooperation evolves. Therefore, here we present a theoretical solution of the cooperation problem among unrelated individuals with imitation capacity. 

\bigskip

\section{The model}

\bigskip

Each individual is represented by a node of a network, whose links determine the interacting individuals. We considered all nodes intrinsically equal and all connections undirected and with equal weight. Each interaction between individuals is modeled by a round of the prisoner's dilemma game. We choose $T=b$, $R=b-c$, $P=0$ and $S=-c$ that allow it to reduce the analisys to a single parameter defined by the benefit-cost ratio $r=b/c$. After one round of the game for each link, the payoff $P_i$ of individual $i$ is defined as follows. If $i$ is a cooperator obtains a payoff $P_{i}=k_{i}^{c}b-k_{i}c$, where $k_i$ is the degree of $i$ and $k_{i}^{c}$ the number of cooperative neighbors. When $i$ is a defector linked to $k_{i}^{c}$ cooperators it gets a Payoff $P_{i}=k_{i}^{c}b$. For a sake of simplicity, we consider individuals without memory or prediction capacity and all strategies of the system are updated simultaneously (synchronous update) following the \emph{weighted democratic update} explained in the next section. A complete update of the system is called a generation. Besides, it is important to state that the conclusions reached through the work are not affected if the strategy update is performed in an asynchronous way. However, it would be of great interest perform a thorough analysis considering this general feature of actual systems. 

\bigskip

We assume that the system growth exponentially by the incorporation of new individuals. These are considered defectors in order to simulate the worst conditions that the cooperative system must to resist. Unless otherwise stated, we do not take into account the genetic relatedness between players and, therefore, the origin of the new nodes is not important for conclusions. Besides, for simplicity, we do not take into account the elimination of nodes but, however, this can be considered without modify conclusions. Between two generations, we leave growth the system a time $\Delta t$. Considering that the system has performed an strategy updates in time $t_0$, the next generation is performed when the system reaches a size $N(t_0+\Delta t)=N(t_0)e^{a\Delta t}=N(t_0)(1+n)$, where $a$ is the rate to which new individuals are introduced to the system. In this way, between two generation the system grows a fraction $n$. It is remarkable that any other growth more slowly than exponential do not modify the conclusion reached. In this situation, it is expected a lower value of $n$ than the one required for exponential growth and, as we will show, it is covered by the results. Otherwise, we consider that individuals are allowed to change their strategy in two different ways, by imitation following the democratic weighted update or by mutation which do not require environmental motivation. In order to take the last into account, we define  $P_m$ as the probability of an individual mutates between two generations. 

\bigskip

Here, we explore the model under three different mechanisms of network growth to perform an exhaustive study of the model. First, we consider the Barab\'asi-Albert model (BAM) \cite{N18}. In this, the system growth from $N_0$ fully connected nodes by the incorporation of new individuals with $L$ links, we consider $N_0=L$. Each new link is attached to an already existing node of the system by preferential attachment, \emph{i.e.} with a probability proportional to the degree of the existing nodes. This model generates networks with a degree distribution governed by a power law $P(k) \sim k^{-\gamma}$ with an exponent $\gamma \simeq 2,9$ in the thermodynamic limit $N \rightarrow \infty$. Second, we explore the model by considering the Model A (MA) \cite{I2}. Starting from a fully linked networks of $N_0=L$, the system grows by the incorporation of new individuals. For each new node, the system introduces $L$ new links. One of them connects the new node with an existing one chosen by preferential attachment. The remaining $L-1$ links are placed between a random chosen node and other one selected by preferential attachment. This procedure also generates scale-free networks where individuals with less than $L$ links are allowed in opposition with BAM. Third, we employ a random network model (RNM). As the previous ones, the system growth from $N_0=L$ fully linked individuals by the incorporation of new ones. Then, the system introduces $L$ new links for each new node in the following way. The new node is linked to a random picked node and the remaining $L-1$ links are placed between any two nodes randomly chosen. This mechanism produces networks with an exponential degree distributions where individuals with less than $L$ links are allowed. It is important to state that in the three network generation models, we avoid the formation of double connections. We choose these models for two reasons. On the one hand, comparing the results of BAM and MA we can determine the importance of the existence of nodes with less than $L$ links, which has shown \cite{I2} to be very important in the formation of cooperative systems without mutations. Besides, we can study the importance of the degree heterogeneity by comparing the results obtained with MA and RNM.

\section{Democratic weighted update}

\bigskip

As we stated in the introduction, we consider that the behavior of any focal player $i$ is affected by all the members of its neighborhood $I$ through the information provided by them, \emph{i.e.} payoff and strategy of each individual in the neighborhood. Here, individuals evaluate their strategy considering this information as follows. First, for a sake of simplicity, we divide the neighborhood in two sets namely $O$ and $S$, where $O$ are the players with different strategy than $i$ and $S$ the remaining ones. Then, we say that the focal player is motivated to change its strategy if two conditions are satisfied. On the one hand, the set $O$ must not be empty. On the other hand, the average payoff $\bar P_O=\sum_{l \in O} P_l/\sum_{l \in O} 1$ of the set $O$ must to be higher than the average payoff $\bar P_S=\sum_{l \in S} P_l /\sum_{l \in S} 1$ of the set $S$, \emph{i.e.} $\bar P_O > \bar P_S$. When these conditions are satisfied the focal player notes that there is a different strategy than its own that, in average, it is doing better. When $O$ and $S$ have a similar number of elements, the average payoffs difference $\bar P_O-\bar P_S$ could be a good measure of which strategy is better. Nevertheless, in the situation where $O$ ($S$) have much more elements that $S$ ($O$), the average payoff of $\bar P_O$ ($\bar P_S$) is statistically more reliable than the other one. In this sense, the abundance of each strategy becomes important information that must be considered. Here, we take this into account by considering that the social influence of each individual over its neighbors is defined by its payoff. In this way, individuals with high payoff have a stronger influence over the behavior of its neighbors than players with low one. Thereby, the total social influence of each strategy over a focal player is defined by the accumulated payoff $P_O=\sum_{l \in O} P_l$ for the strategy $O$ and $P_S=\sum_{l \in S} P_l$ for the strategy $S$. Thus, the effective social influence over the focal player is defined by $P_O-P_S$. In this way, when $P_O>P_S$ ($P_O<P_S$) the social influence favors the strategy of the set $O$ ($S$) and when $P_O=P_S$ the social pressure does not have effect over $i$. Thus, we consider that a motivated player changes its strategy with a probability $w$ that increases with the effective social influence $P_O-P_S$. We perform this by introducing the Fermi function (ref) as follow: 

\bigskip

\begin{equation}
w=\frac{1}{1+e^{-\beta (P_O-P_S)}}\text{ ,}
\end{equation}
  
\bigskip

where $\beta$ is the intensity of social influence over the focal player. When $\beta \rightarrow 0$ the abundance of each strategy becomes neutral and the individual $i$ changes its strategy with a probability $w \rightarrow 1/2$. Otherwise, when $\beta \rightarrow \infty$ the focal player changes its strategy with $w \rightarrow 1$ ($w \rightarrow 0$) when $P_O>P_S$ ($P_O<P_S$) and, therefore, the influence of the neighborhood becomes determinant in the behavior of $i$. It is interesting to note that if $\beta=0$ or $P_O=P_S$ the motivated focal player preserves or changes its strategy with equal probability since $w=1/2$. This situation can be interpreted as the individual $i$ trusts in its last choice as much as in the best average strategy of its neighborhood in the last round. Therefore, the focal player valuates more its own information than the provided by a neighbor. Otherwise, when $\beta \rightarrow \infty$ and $P_O \neq P_S$, the focal player considers equally important the information provided for each member of its neighborhood. Lastly, for intermediate values of $\beta$ and $P_O \neq P_S$, the focal player $i$ considers its information more important than the provided by a neighbor at the same time that the social influence is considered. 

\bigskip

Besides, since $w \neq 0$ for all $\beta$ and $P_O-P_S$, any motivated player is allowed to change its strategy even when $P_O<P_S$. This is important because allow to individuals overcome the social influence taking a strategy in opposition to the wisdom or pressure of the group. Here, the motivation of a focal player to change its strategy has been defined in a deterministic way. However, the strategy update can be generalized by considering a motivation probability $m$ of change strategy that increases with $\bar P_O -\bar P_S$. In particular, this can be newly performed through the Fermi function defined by $m=1/(1+e^{-\alpha (\bar P_O-\bar P_S)})$, where $\alpha$ is the intensity of natural selection. In this way, when in a neighborhood there are two strategies, the probability that the focal player changes its strategy is defined by $P_{S \rightarrow O}=m \times w$, where the previous situation is recovered for $\alpha \rightarrow \infty$. It is interesting to note that when $\beta=0$, the \emph{democratic weighted update} is reduced to the strategy update proposed in \cite{S1}. However, although it would be interesting to explore thoroughly the influence of $\alpha$ in the results of the model, we just consider the case where $\alpha \rightarrow \infty$. However, we have checked that the conclusions reached through the work are unaltered considering any $\alpha > 0.01$. Therefore, the conclusions are robust to this consideration. 

\bigskip

\section{The formation of highly cooperative systems without mutations}

\bigskip

To perform a clear analysis of the model, we start exploring it for $P_m=0$. To this, we look for the critical $r_c$ required to maintain a stable and high level of cooperation when the system grows from an initial structure of $N_i$ cooperators by the incorporation of new defectors. Through the paper we consider $N_i=1000$ in order to ensure a proper development of the topological structure of the network under consideration. Nevertheless, we can justify this initial cooperative structure by extending the results of \cite{I2}, where it has been introduced considerations that allow the formation of this initial structure of cooperators. In fig. $1$ we show the fraction of cooperators $\bar c$ as a function of the benefit-cost ratio $r$ for the three growing mechanisms explored and different values of the parameters $L$, $\beta$ and $n$. These results have been obtained averaging $100$ different realizations of the model, which correspond to simulations where the system has grown to $N=10^4$. The fraction of cooperators $\bar c$ of each realization has been obtained averaging the level of cooperation reached by the system after each generation for $N>9\times 10^3$.

\bigskip

In fig. 1(a) we can see the results obtained by the three growing mechanisms explored, \emph{i.e.} BAM, MA and RNM. In all cases we observe a phase transition from a non-cooperative state to a highly cooperative one with some critical benefit-cost ratio $r_c$. In particular, it is remarkable the very low $r_c$ required for cooperation to evolve irrespective of the underlying network formation mechanism. Also, as it is expected \cite{N3,N4,N5}, we observe by comparing the results of RNM and MA that $r_c$ decreases when the degree heterogeneity increases. Nevertheless, the benefit for the evolution of cooperation produced by increasing the degree heterogeneity of the system is much less significant than the one of already formed systems \cite{R1,R2}. Otherwise, we shown again \cite{I2} that MA requires a lower value of $r_c$ than BAM for cooperation to evolve. Besides, it is very interesting to note that RNM presents a $r_c$ lower than the one of BAM even when RNM is much less heterogeneous than BAM. This remarkable result without precedent in the literature shows newly \cite{I2} the great importance of avoid the introduction of new individuals with many connection performed simultaneously. When the systems is highly cooperative, the capacity of a defector to exploit the system is approximately proportional to its degree $k$. Therefore, when the new individuals are introduced to the system with many connections ($L$) simultaneously instead one at a time, the system requires a higher $r_c$ to overcome the incorporation of these strong new defectors and, therefore, the required $r_c$ for cooperation to evolve increases. 

\bigskip

\begin{figure}[!hbt] \centering
\includegraphics[angle=0,scale=0.6]{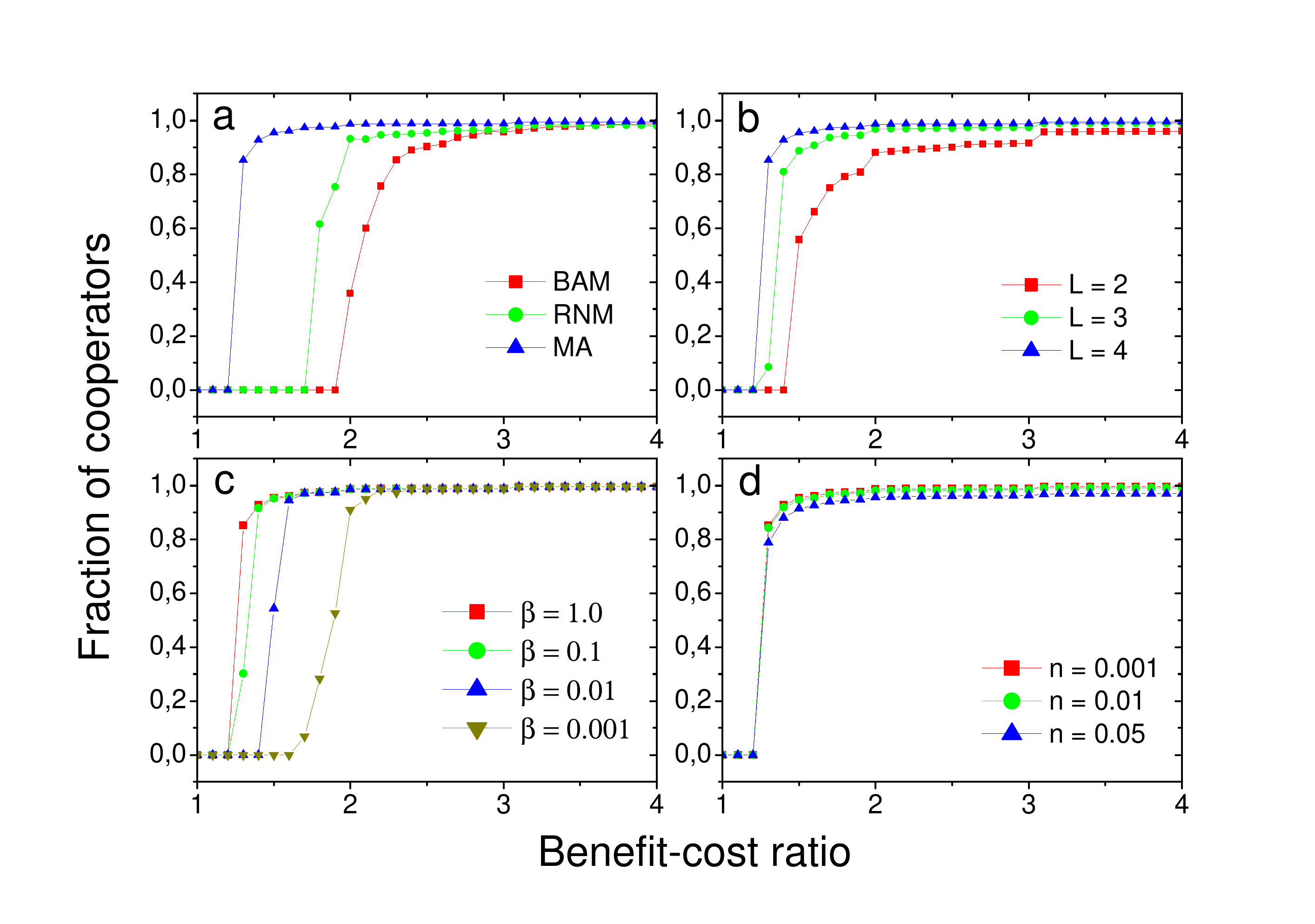}
\caption{The model without mutations. The fraction of cooperators as a function of the benefit-cost ratio for different network growth and parameters of the model. (a) The model over different growing mechanisms. (b) The influence of the average degree $\bar k \simeq 2L$. (c) The dependence with the intensity of social pressure $\beta$. (d) Exploring the model over different values of $n$. The parameters not specified in each figure correspond to $L=4$, $\beta=1.0$, $n=0.001$ and using MA as growing mechanism.}
\end{figure}

\bigskip

In fig. 1(b) we explore the influence of the average degree $\bar k \simeq 2L$ in the required $r_c$ for cooperation to evolve. As well as in \cite{I2}, the critical $r_c$ decreases approaching to the theoretical minimum $r_c=1$ when $\bar k$ increases. When the systems is highly cooperative the payoff of individuals is approximately proportional to its degree. Besides, considering that cooperators inhabit the most linked parts of the system and defectors the lowest ones \cite{I1,I2}, the average payoff difference between defectors and cooperators increases with $\bar k$ and, therefore, $r_c$ decreases with $\bar k$. Nevertheless, it is important to note that for any $L$ the system presents a very low $r_c$. Besides, we observe the existence of other transitions that increases slightly the fraction of cooperator into the system. In particular, these transitions are clear in $r=2$ and $r=3$ for $L=2$, and also in $r=2$ for $L=3$. As it has been shown \cite{I2}, when a defector is connected only with a cooperator who in turn is connected with $k$ other cooperators, for $r>(k+1)/(k-1)$ the cooperator have a higher payoff than the defector and, therefore, the cooperator is favored by natural selection. In particular, the cases $k=2$ and $k=3$ correspond to the transitions $r=3$ and $r=2$ respectively. When the average degree of the system increases, the probability of a defector to be connected with a cooperator with few connection decreases and, therefore, these transitions become less evident. 

\bigskip

In fig. 1(c) we can see how the model behave for different values of the intensity of social influence $\beta$. In particular, it shown that according as $\beta$ increases the critical $r_c$ decreases. In this way, we show that the social influence have a positive effect for the evolution of cooperation when $P_m=0$. However, the cooperation evolves with a very low $r_c$ irrespective of $\beta$. Besides, it is interesting to note that $r_c$ goes to the one showed in \cite{I2} when $\beta \rightarrow 0$. This is important because shows that previous conclusions are robust when it is considered the average payoff of each strategy instead a pairwise comparison. Also, since $\beta$ does not modify drastically $r_c$ we conclude that the social influence is important but not determinant in the formation of cooperative systems without mutations. Otherwise, in fig. 1(d) we show results of the model for different values of $n$. As we can see, $r_c$ is independent of $n\leq 0.05$. The only consequence of $n$ is a diminution in $\bar c$ when $n$ increases. This is expected since $n$ determines the number of new defectors between two generations. Finally, it is important to state, that the initial structure of $N_i$ cooperators can be justified by extending the results of \cite{I2} since here we largely shown that the required $r_c$ is always improved by the social influence. In this way, we have justified the formation of highly cooperative systems of any size when $P_m=0$.

\bigskip

At this point, we could be interested in the consequences of the social influence in already formed structured populations. Here, for space reasons, we limit to comment that the $r_c$ required for cooperation to evolve is deeply worsened, in particular when $L$ increases. Therefore, we are newly showing the great importance of the growing process for the evolution of cooperation in systems whose individuals have imitation capacity and the mutation probability is low enough to be neglected. Under these conditions, the cooperators inhabit the most linked parts of the system and defectors the lowest ones. Since this arrangement of cooperators and defectors into the system is the optimal for cooperation, the required $r$ for cooperation to evolve is drastically reduced with respect to one of already formed systems. In particular, we have broadly shown that the system requires a very low $r_c$ independently of the parameters of the model shown, the growing mechanisms and the strategy update considered. Besides, we have checked that equivalent conclusion can be reached for any traditional game, \emph{i.e.} the snow-drift game, the stag-hunt game and public good games. Therefore, given the great robustness of conclusions, we propose the growing process as an universal mechanism for the evolution of cooperation when $P_m=0$ and the individuals of the system imitators. 

\bigskip

\section{The highly cooperative system overcoming mutants}

\bigskip

In the last section, we have shown that the required $r_c$ for cooperation to evolve when $P_m=0$ is very low independently of the remaining parameters of the model and the growing mechanism considered. However, in order to show that the model is a complete solution of the cooperation problem formulated in \cite{I1}, it is required that cooperation prevails into the system even when $P_m \neq 0$. Since the impact of mutant defectors over the system is very sensitive to the area of the network where they appear, it is very important to explore the emergence of mutants considering a lot of generations to cover a great number of possible configurations of defectors into the system. To this, we address the problem considering the apparition of mutants between generations but not the incorporation of new defectors. In this way, we study the evolution of the system when mutants appear in a system with fixed size $N$ and considering a fully cooperative system as initial condition. We choose this way to explore the model because it allows to consider efficiently many generations since otherwise the system size $N$ required to reach the same number of generations becomes very large, and therefore, computationally inefficient. Besides, this procedure helps to focus the attention just over the mutants since they are the only source of defectors. However, it is important to state, as we will show (see fig. $3$), that almost equivalent results are obtained when both processes are considered simultaneously for $N_i \geq 1000$. Then, we will extend results to systems smaller than $N=1000$ in order to show that the model allows cooperation to evolve under very low conditions for any $N$. The results obtained considering a fixed system size have been obtained averaging $100$ different realizations of the model, where fully cooperative systems of $N=10^4$ individuals as initial condition have been considered. The fraction of cooperators of each realization has been obtained averaging the level of cooperation over $10^3$ generations after a transient of $10^4$ generations.  

\bigskip

\begin{figure}[!hbt] \centering
\includegraphics[angle=0,scale=0.60]{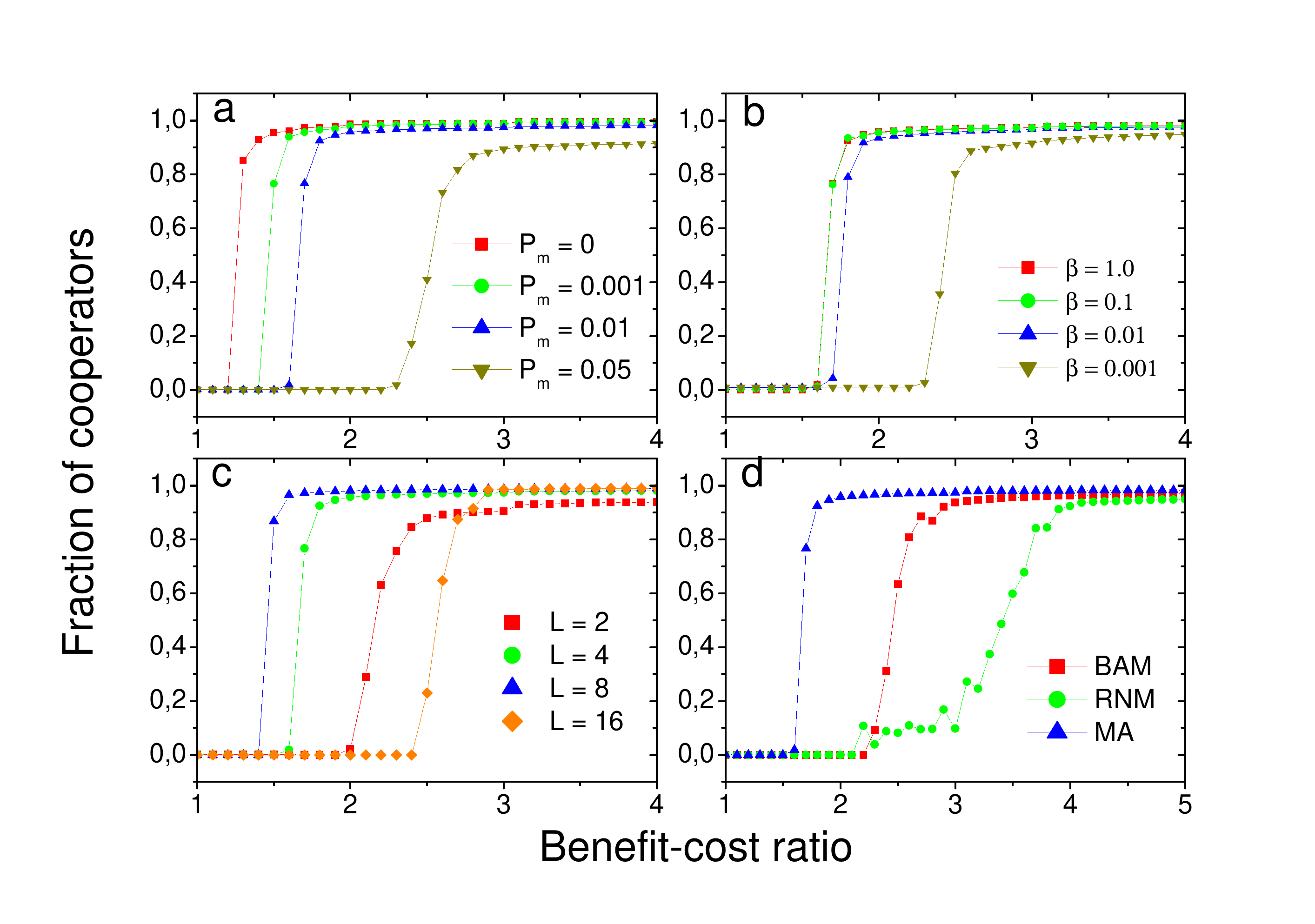}
\caption{The emergence of mutants in highly cooperative systems. The fraction of cooperators as a function of the benefit-cost ratio for different network growth and parameters of the model. (a) The model for different values of the mutation probability $P_m$. (b) The dependence with the intensity of social influence $\beta$. (C) The effect of the average degree $\bar k \simeq 2L$. (c) The model over different growing mechanisms. The parameters not specified in each figure correspond to $L=4$, $\beta=1.0$, $P_m=0.01$ and using MA as growing mechanism.}
\end{figure}

\bigskip

In fig. $2$ we show the fraction of cooperators into the system as a function of the benefit-cost ratio $r$ for the three growing mechanisms considered and many values of the parameters of the model $L$, $\beta$, $n$ and $P_m$. In particular, fig. $2$ (a) shows results of the model for different values of $P_m$. As we can see, the system presents a phase transition from a non-cooperative to a cooperative one for some $r_c$. In particular, we observe an increment of $r_c$ with $P_m$. However, it is remarkable that the system present a low $r_c$ even for the very high mutation rate $P_m=0.05$. In this way, we shown that the social influence allows to overcome the apparition of mutant defectors in highly cooperative systems. It is important to state, that for some $P_m>0.05$, as it is expected, the system cannot reach a high and stable value of cooperation regardless of the value of $r$ given the great noise in the behavior of individuals. Otherwise, in fig. $2$ (b) we show results of the model for different values of the social influence intensity $\beta$. We observe that $r_c$ is not strongly dependent for $\beta>0.01$ but it increases fast for $\beta<0.01$. In particular, we have checked that for $\beta=0$ is not possible to reach a stable value of cooperation for any $r$. Nevertheless, the cooperation evolves with a very low $r_c$ for all $\beta$ considered. Therefore, the social influence is a very strong mechanism to preserve cooperation in highly cooperative systems despite the emergence of mutant defectors.

\bigskip

\begin{figure}[!hbt] \centering
\includegraphics[angle=0,scale=0.40]{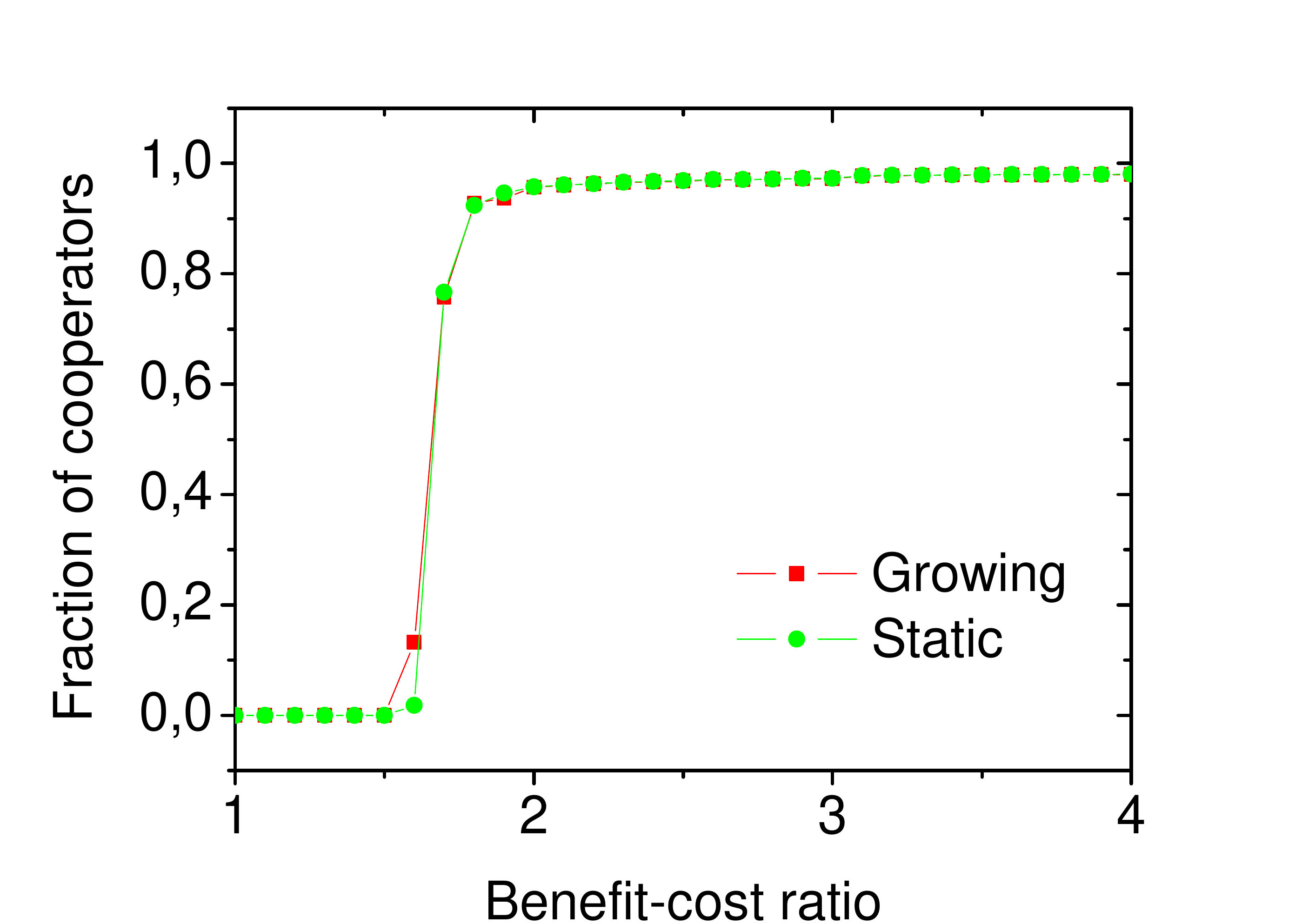}
\caption{The equivalence between growing systems and already formed highly cooperative ones. In both cases we employ MA for the network formation, $P_m=0.01$, $L=4$ and $\beta=1$. For the growing results we choose $n=0.001$ and a final system size $N=10^4$. Otherwise, the results of the static case have been obtained for $N=10^4$.}
\end{figure}

\bigskip

In fig. $2$ (c) we show the behavior of $\bar c(r)$ for different values of $L$. Surprisingly, the model presents an optimal $\bar k$ where $r_c$ is minimal. This is a very interesting result that could help to understand why actual networks \cite{N20} has evolved to the everage $\bar k$ observed. In particular, it would be of great interest to develop a network generation model where the average degree of the system is determined by coevolutionary dynamics \cite{C1} instead of being exogenously introduced. Nevertheless, it is very important to note that the system presents a low $r_c$ for all the $\bar k$ considered and, therefore, we are able to conclude that the model is very robust to $\bar k$. Moreover, in fig. $2$ (d) we show results for the three growing mechanisms considered. We observed that cooperation requires a very low $r_c$ to evolve irrespective of the underlying growing mechanism and, therefore, the model is very robust to this consideration. Besides, it is remarkable that MA and even RNM reduce $r_c$ with respect to the one required with BAM. In this way, we are showing that the existence of individuals with less than $L$ links is also very important to overcome the emergence of mutant defectors. This feature is generally neglected in agent-based models but it could have also important implications. Thus, we hope with this result motivates further researches considering MA as the skeleton of the system. Otherwise, we observe, for the RNM results, a smoother phase transition than the one obtained with MA. In particular, the region where $\bar c$ reaches intermediate values correspond to the situation where some realizations finish with a high level of cooperation and other ones near zero cooperation. This very important result shows that the life expectancy of the cooperative system increases with $r$. In this way, by considering an greater number of generations is expected smoother transitions for any growing mechanism considered. Besides, by comparing the RNM and MA results, we observe a higher life expectancy as well as a lower $r_c$ when the degree heterogeneity of the system increases. Anyway, we have considered a great number of generations and realizations for results and, therefore, the model generates cooperative systems with a great life expentancy. Otherwise, for completeness and as it is expected, we have corroborated that cooperation is extinguished after some generations in fully connected systems for any $r$ considered. In this way, the structure of the system is an essential feature for cooperation to evolve. However, it is remarkable that even for the random growth mechanism cooperation is able to evolve for a very low $r_c$. Finally, in fig. $3$ labeled as growing, we show results obtained considering the growing process simultaneously with the apparition of mutants. Besides, labeled as static, we show the results obtained with the same parameters values but without consider the growing process. As we can see, both results are almost equivalents and, thus, we confirm that the conclusions can be extended to the case where the growing process is considered simultaneously with the emergence of mutants. 

\bigskip

\begin{figure}[!hbt] \centering
\includegraphics[angle=0,scale=0.40]{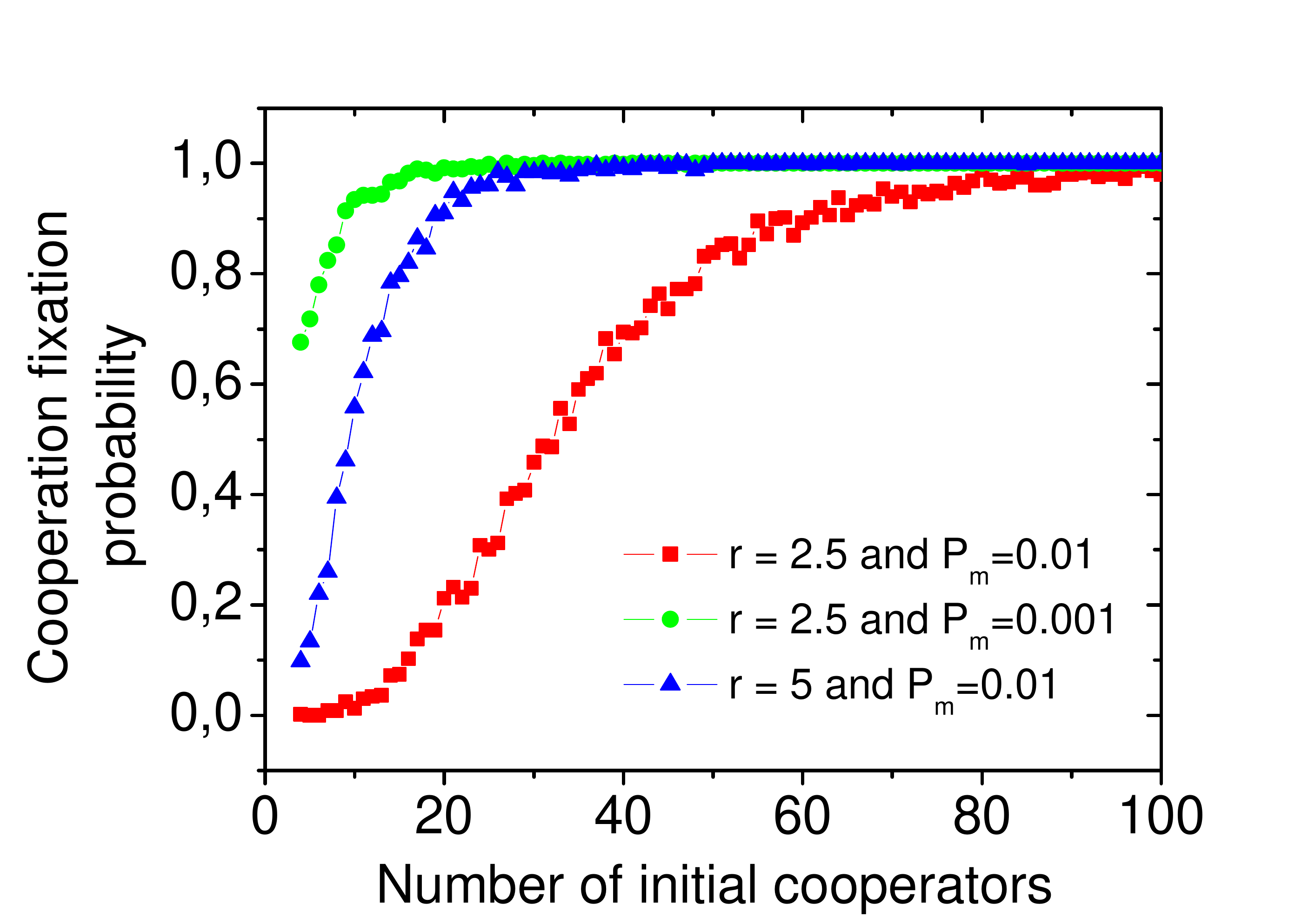}
\caption{The cooperative seed. Cooperation fixation probability as a function of the number of initial cooperators. The model parameters do not specified in the figure correspond to $n=0.01$, $L=4$ and $\beta=1$. We used MA for the network formation.}
\end{figure}

\bigskip

We have widely shown that the social influence allows to highly cooperative systems resist the apparition of mutants for all the parameters considered whenever $N \geq 10^3$. In order to extend these conclusions to systems with $N<10^3$ we consider the cooperation fixation probability $P_{f}(N_{i})$ defined as the probability that a system of $N_{i}$ cooperators continues being highly cooperative when it grows by the incorporation of defectors and considering mutations. To obtain $P_{f}$ we select a value $r>r_{c}$ and we perform $M$ simulations starting from a system of $N_{i}$ cooperators. Then, we compute the number $M_{c}$ of systems that reach $N=10^{4}$ with a fraction of cooperators $\bar{c}>1/2$. Finally, we compute $P_{f}$ for each $N_{i}$ as $M_{c}/M$, we consider $M=500$. As it can be seen in fig. $4$ for all curves, $P_{f}$ grows steeply and reaches the value $1$ for some systems size. From this structure, called \emph{Cooperative Seed} (CS) \cite{I1}, the cooperation is stable despite the system grows by incorporating defectors and the emergence of mutants defectors. However, it is important to note that the instability of the cooperation into the system is generated by the mutant defectors but not by the new individuals given the results of the previous section. We observe that the minimal system size required for cooperation to be stable is reduced when $r$ increases. In this way, $r_c$ decreases quickly with $N$ when the system is small but, as soon as the degree distribution of the system is well developed, $r_c$ becomes approximately size independent. It is interesting to note that the existence of the \emph{cooperative seed} is expected since the system starts to grow from a fully connected system, which cannot support mutant defectors. In this way, the system need to develop payoff heterogeneities from degree ones in order to ensure the stability of the cooperation. However, considering fig. 2 (d) and that $r_c \neq r_c(N)$ from some system size, we expect the existence of a optimal payoff heterogeneity where cooperation is ensured for a low $r_c$ at the same time that these heterogeneities minimal. This is of great interest for the search of a more egalitarian society. Nevertheless, for reasons of space, we leave these explorations for further works. Otherwise, we observe that the size $N_c$ of the CS decreases with $P_m$ for a fixed $n$. However, if $P_m$ and $n$ are not fixed, we have checked that $N_c$ increases with $P_m/n$. In this way, we have shown that combining the growing process, social influence and individuals with imitation capacity, cooperation evolves under very low conditions when $N>N_c$. In particular, when the system is small the cooperation becomes unstable but with a high probability to overcome the critical system size $N_c$. Thus, it would be of great interest to consider other features that allow reducing $N_c$. In particular, if the new individuals come from the reproduction of already existing ones, it is expected high genetic relatedness between linked individuals when the system is small. Then, it could be considered a $P_m$ that decreases with the genetic relatedness among individuals. In this way, $P_m$ would be low when the systems is small and the size of the CS reduced. This assumption can be justified considering that mutations between genically related individuals supposed that the survival probability of the offspring decreases \cite{I2}. Therefore, it is expected that individuals with low mutation rate between related individuals evolve. Anyway, it would be of great interest performs a thorough analysis looking for the optimal $P_m=P_m(N)$ that ensures the best performance of players.

\bigskip

\begin{figure}[!hbt] \centering
\includegraphics[angle=0,scale=0.40]{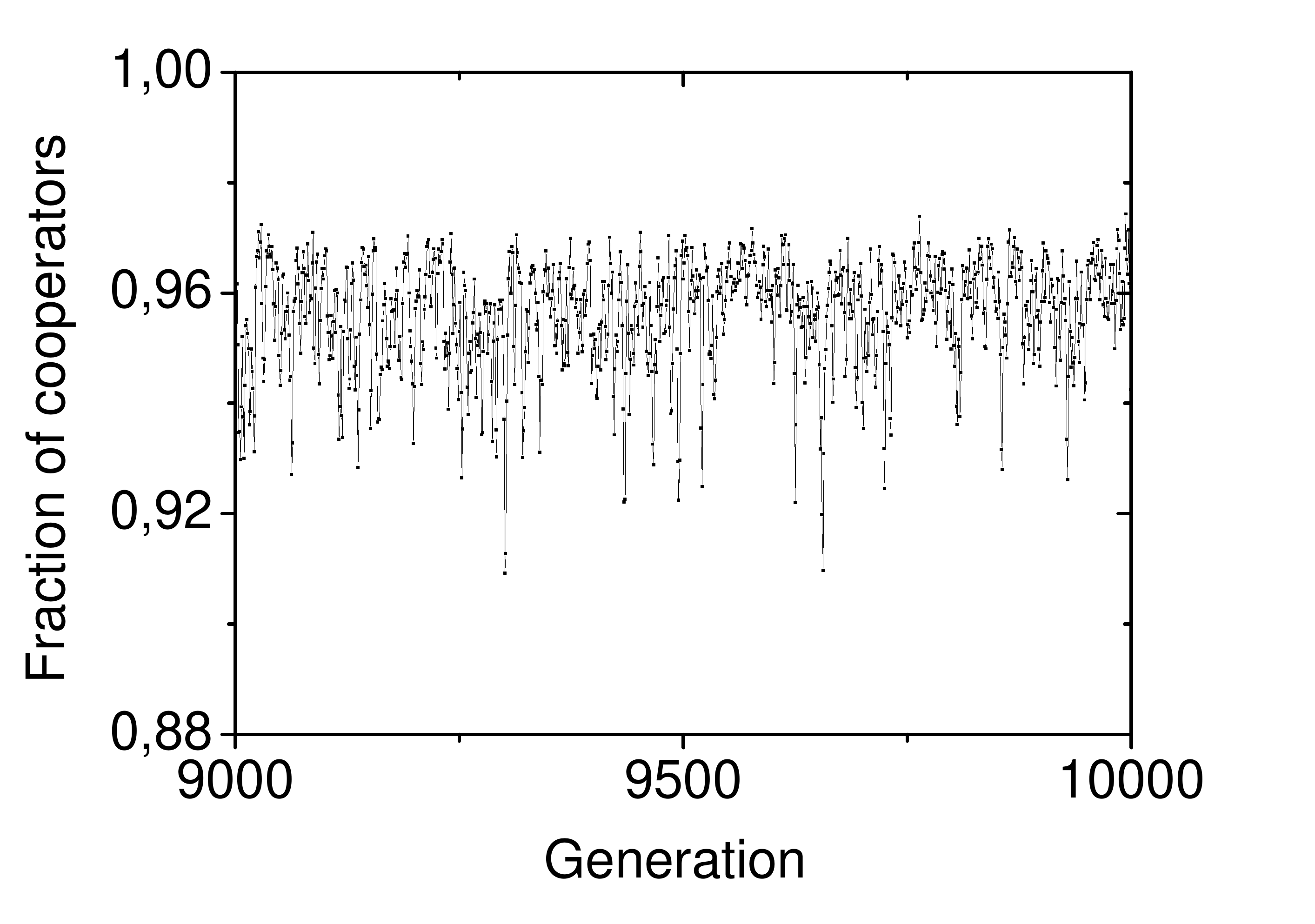}
\caption{The fraction of cooperation of a single realization as a function of the generations of the system. The results have been obtained for $L=4$, $\beta=1.0$, $r=2.0$ and using MA for the network generation.}
\end{figure}

\bigskip

Now, in order to better understand the underlying mechanism through which the social influence allow overcoming the emergence of mutant defectors, in fig. $5$ we show the fraction of cooperation as a function of the generations of a single realization of the model. We observe that $\bar c$ presents some irregular oscillations around the mean value $\bar c \simeq 0,96$. These are produced by the emergence of mutant defectors strong enough to spread their strategy into the system by imitation. After the apparition of these mutants, the fraction of cooperation decreases but, as we can see, the system is able to overcome these invasions by restoring the level of cooperation. It is interesting to note that the defection invasions are of short range since the system always preserves a high level of cooperation. However, it is important to state that increasing $L$ the defection invasions become deeper as it is expected considering fig. $2$ c. This could be particularly important to develop the previous suggested network generation model where the average degree of the system is determined by coevolutionary dynamics. Otherwise, the invasions are short lived because the cooperation level is restored after few generations. In particular, we have checked that the range of the invasions depends of the degree of the mutant defector as well as the number of mutant defectors that appear into the systems between generations. In this sense, the deepest invasion observed is produced by several highly connected mutants instead by just one. Also, it is very important to note that if many highly linked individuals mutate simultaneously the system can die for any $r>r_c$. Nevertheless, this occurs with very low probability $P_d$ as we extensively shown in fig. $2$ given the great number of generations and realizations considered for the results. Even so, the death of the organism is ensured since $P_d \neq 0$ and, as we stated in the introduction, it is essential for the new organism acquires reproduction capacity in order to avoid the extinction.

\bigskip 

\begin{figure}[!hbt] \centering
\includegraphics[angle=0,scale=0.40]{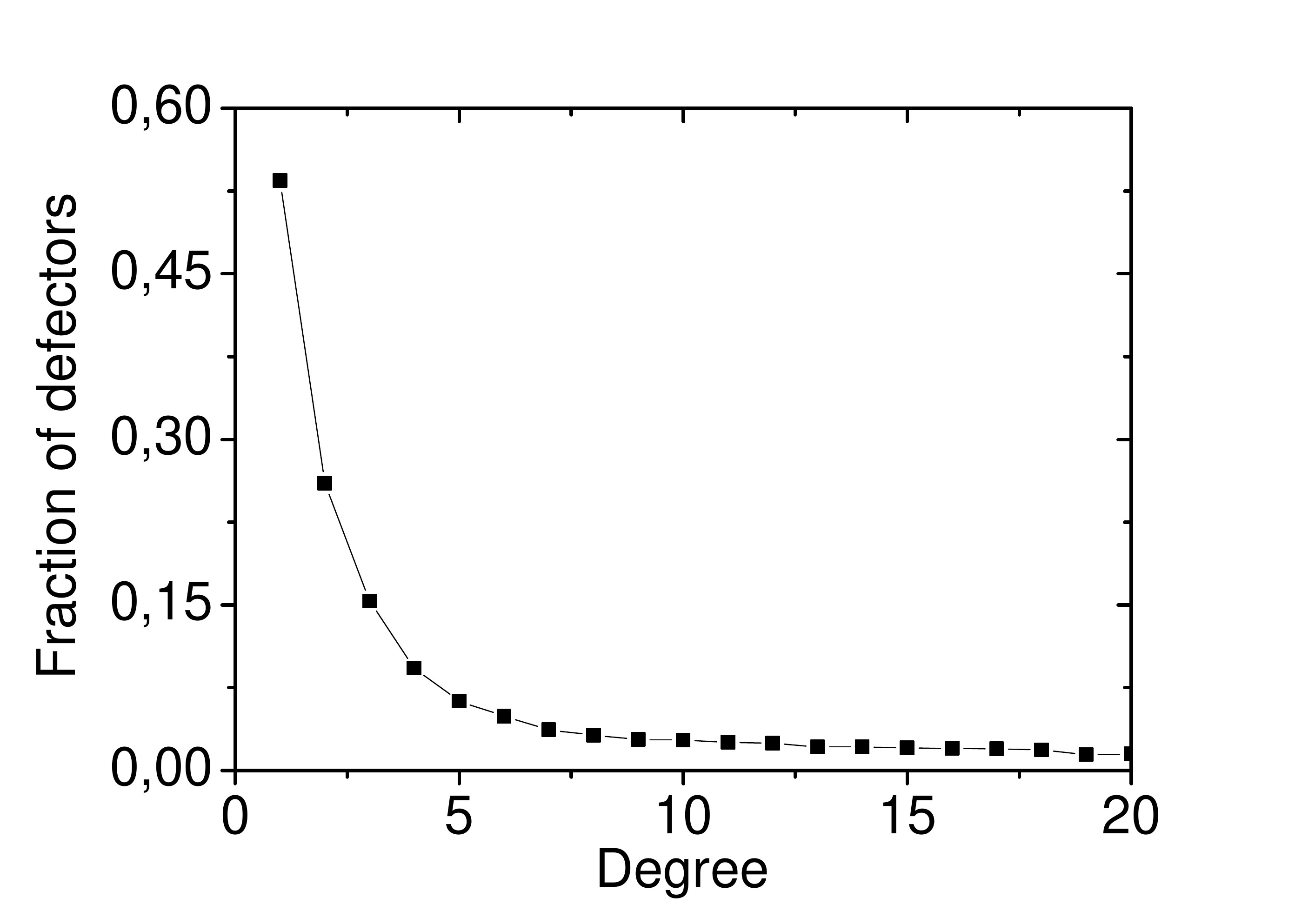}
\caption{Fraction of defectors as a function of the degree of nodes. The results have been obtained for $L=4$, $\beta=1.0$, $r=2.0$, $P_m=0.01$ and using MA for the network generation.}
\end{figure}

\bigskip

In fig. $6$ we show the fraction of defectors in function of the degree of the nodes. These results have been obtained averaging over $9 \times 10^3$ generations after a transient of $10^3$ one and $10$ different networks. As we can see, the fraction of defectors decreases fast with the degree of individuals approaching to $0,01$ which correspond to $P_m$. In this way, defectors inhabit preferably the low linked individuals of the system and cooperators the highly connected ones. Considering that the result have been obtained without consider the incorporation of new individuals, we conclude that the invasions produced by the emergence of defectors spread to the low linked individuals when the social influence is considered. In this way, it is very important to the cooperative system the existence of individuals with low degree in order to reduce the explotation capacity of defectors into the system. Therefore, this model behavior explain why MA and RNM present a $r_c$ lower than BAM. Otherwise, in order to understand the how the social influence allows to overcome the apparition of mutant defectors producing this distribution of defectors into the system, we analyze the mechanism by which a mutant defector newly becomes cooperator by social influence in a heterogeneous highly cooperative system. In particular, in order to explore the worst situation possible, we consider the case where a hub of a fully cooperative system mutates. When this occurs, the mutant increases its payoff next round since all its neighbors are cooperators. However, in the next generation its neighbors have a new possibility previously nonexistent, now they could change by imitating the strategy of the mutant. Since this is a hub, it is expected $\bar P_O>\bar P_S$ in the neighborhoods of its neighbors and, therefore, all of them are motivate players. At this point it is important to note that the accumulated payoffs $P_S^j$ and $P_O^j$ of a neighbor $j$ of the mutant can be expressed as $P_S^j=(k_j-1)\bar P_S^j$ and $P_O^j=\bar P_O^j$ in the first generation after the mutation respectively. In this way, it is expected $P_O^j-P_S^j$ smaller in average for high connected neighbors than for low linked ones. Therefore, the probability of a neighbor to change its strategy next generation decreases in average with its degree. In particular, considering $\beta \rightarrow \infty$ and $r$ high enough, the low linked neighbors adopt defection by imitation and the highly linked ones preserve cooperation as strategy. In this situation, the mutant reduces its payoff next generation since now it have lot of new defector neighbors. Besides, its average payoff $\bar P_S$ is reduced since its defectors neighbors are poorly connected to the systems and, therefore, with a low payoff. However, $\bar P_O$ of the mutant increases because all the neighbor cooperators are the highly linked ones. Thus, the mutant becomes in a motivated player since in its neighborhood $\bar P_O>\bar P_S$ if $r$ if high enough. Therefore, in the next generations the mutant becomes newly cooperator. In this way, after the emergence of a mutant in a fully cooperative system, we have shown that defection spread preferably to the low linked nodes.

\bigskip

In this way, through this section we have widely shown that a highly cooperative system formed by a growing process is broadly able to overcome the emergence of mutants by considering the social influence. Nevertheless, it is natural to ask about the outcome of the model if the abundance of strategies is considered through the learning activity \cite{S2} instead by social influence. Although it would be interesting a thorough analysis of this, we perform a brief one in appendix A in order to newly show the great importance of take into account the abundance of strategies as well as the advantages of the social influence.

\bigskip

\subsection{Testing the model in human experiments}

\bigskip

We believe of great interest to perform human experiments considering the growing process of the system given that it has theoretically shown \cite{I1,I2} to generate a favorable environment for cooperation to evolve irrespective of the underlying strategy update rule and parameters of the model considered. To this kind of experience, the \emph{cooperative seed} becomes particularly important due the great initial mutation rate reported in human experiments \cite{E1}. Besides, it would be of great interest to test if the human being follows a strategy updates similar to the \emph{democratic weighted update}. In this sense, a very important step towards the develop of a theoretical model of the human society is to understand how the information provided by the neighborhood affects the behavior of the human beings. Taking this as motivation, it has been performed several great human experiments \cite{E1,E2,E3} where this problem has begun to be addressed. Although these experiments have shown several important consequences, here we just highlight those that we consider more relevant. In \cite{E1} has been shown that the human mutation rate is much higher than typically assumed in theoretical models. In particular, it has been reported a very high initial mutation rate which decays exponentially in time. Otherwise, in \cite{E2} have been shown that individuals do not follow the well-known imitation to the best rule \cite{N1}. Lastly, in \cite{E3} has been shown that degree heterogeneity do not promote cooperation when it is considered the payoff of individuals normalized by their degree. When these conclusions are taken into account, a great number of theoretical predictions \cite{R1,R2}, which are very susceptible to the strategy update rule considered, are no longer valid to fully explain human cooperation. Nevertheless, it is important to note that the experimental conditions considered in these works \cite{E1,E2,E3} are very different to the everyday human reality and, therefore, the results obtained can still not be conclusive. In particular,  the experiments have been performed considering the payoff of individuals normalized by their degree instead of considers the accumulated one. This could have consequences as important as it does in theoretical models \cite{N11,I2} and, therefore, it would be of great interest to consider this feature which clearly is present in the actual human society. 

\bigskip

\section{Conclusions and new perspectives}

\bigskip

In this paper we have addressed the cooperation problem over structured populations considering the formulation performed in \cite{I1} and the prisoner's dilemma game as metaphor of the social interactions. We have introduced a new strategy update called \emph{democratic weighted update} considering that the individuals influence the behavior of their neighbors. In particular, we have considered that the capacity of individuals to influence others is given by their wisdom which has been defined as proportional to their payoff. Besides, when in a neighborhood there are cooperators and defectors, the focal player is influenced contradictorily by the wisdom of individuals and, therefore, the effective social influence is defined by the difference between the total wisdom of each strategy in the neighborhood. We have extensively explored the model for a wide range of the parameters and several growing mechanism showing, in any case, very low conditions for cooperation to evolve. In this way we conclude that considering the growing process of the system, the social influence and individuals with imitation capacity cooperation evolves and, therefore, we have shown a complete theoretical solution of the cooperation problem among unrelated individuals with imitation capacity. Besides, we have extended the conclusions to other ways to consider the abundance of strategies in the neighborhood of individuals showing its great importance to overcome the emergence of mutant defectors in highly cooperative systems. Thus, we hope to have taken a step towards a theoretical model of the human society. 

\bigskip

We believe that the model has very important implications in further theoretical researches. In particular, the model allows assuming a highly cooperative system as initial condition in agent-based models which has never been taken into account. Here, we have considered that individuals are allowed to imitate cooperation or defection. However, it is possible to consider other features of human being that spread into society by imitation such as rumors, culture, opinions, thoughts, ideologies, etc. In this sense, it would be of great interest understand how the success of individuals, mainly determined by the underlying cooperative system, affect the propagation dynamics of this kind of information. In particular, it could be useful to understand how the knowledge of the human society increases in time. To this, we believe a good first hypothesis to consider the existence of the truth and that the new knowledge emerges from the mutation of an individual previous one. In this way, the knowledge spreads by imitation favoring those which are closer to the truth.

\bigskip



\section{Acknowledgements}

\bigskip

I acknowledge financial support from Generalitat de Catalunya under grant 2009-SGR-00164, Ministerio de Econom\'ia y Competitividad under grant FIS2012-32334 and PhD fellowship 2011FIB787. I am grateful to the Statistical Physics Group of the Autonomous University of Barcelona. I would like to thank David Jou and Vicen\c c M\'{e}ndez for their unconditional support and assistance with the manuscript. In particular, I am deeply grateful to Carolina Perez Mora for the extensive discussions about cooperation.

\appendix

\section{Introducing the abundance of strategies through the learning activity}

\bigskip 

\begin{figure}[!hbt] \centering
\includegraphics[angle=0,scale=0.40]{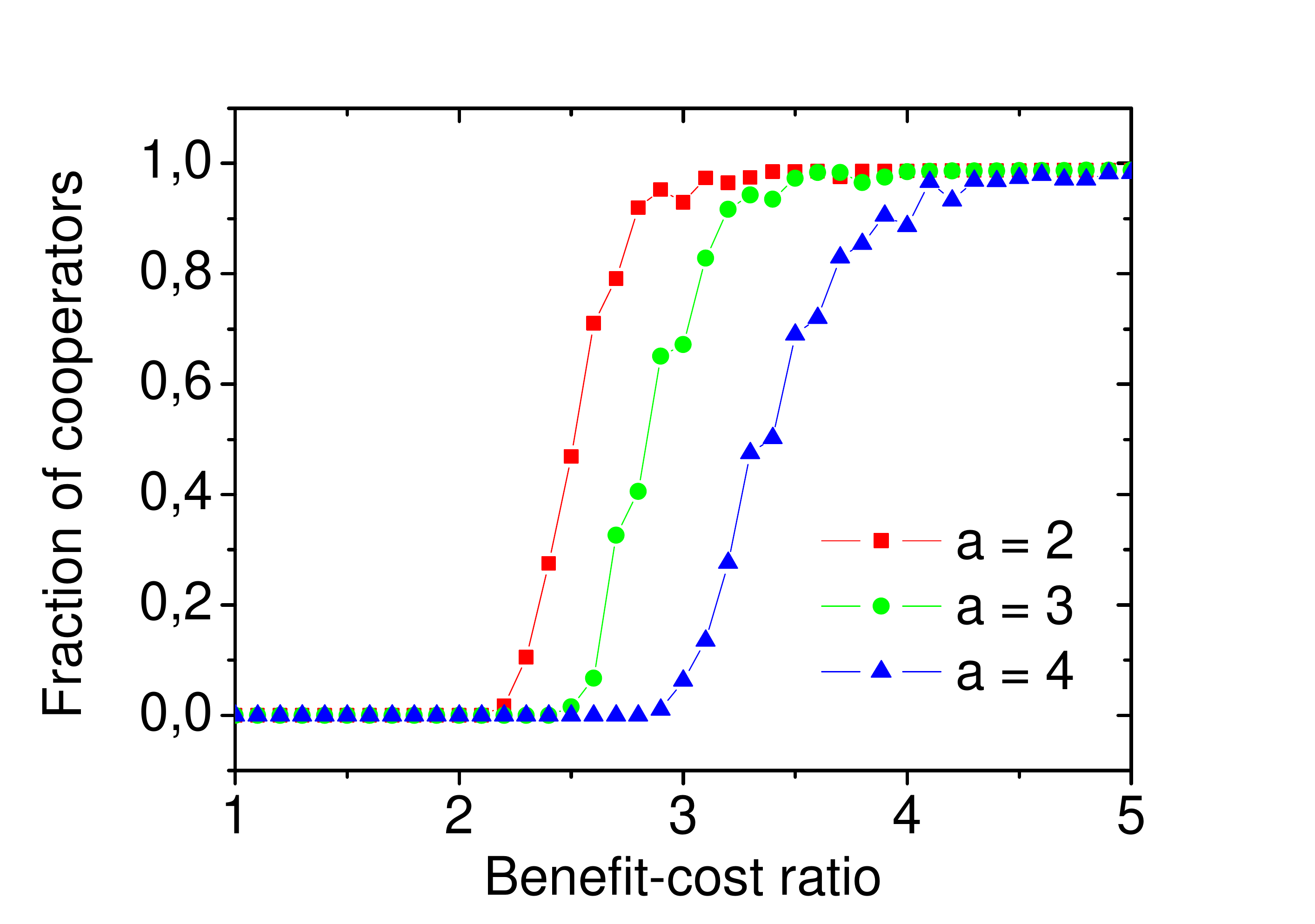}
\caption{Considering the abundance of strategies through the learning activity of individuals. Fraction of cooperators as a function of the benefit-cost ratio for different values of $a$. The results have been obtained for $L=4$, $r=2.0$, $P_m=0.01$ and using MA for the network generation.}
\end{figure}

\bigskip

As we stated in the introduction, the abundance of strategies in the neighborhood of the focal players has been recently \cite{S2} introduced through the learning activity of individuals. In particular, they have defined the learning activity $L_i$ of individual $i$ as $L_i =(\sum_{l \in O}1/k_i)^a$, where $a$ determines how seriously the abundance of strategies affects the behavior of individuals and the degree $k_i$ of the focal player $i$ ensures a proper normalization of the transition probability $w$. In this appendix, we address the question about what happen with the outcome of the model if the behavior of the focal players is affected by the abundance of each strategy through the learning activity $L_i$ instead through the effective social influence $P_O-P_S$. In fig. $7$ we show the fraction of cooperation into the system as a function of $r$ for different values of $a$. These results have been obtained following the same procedure that for the ones of fig. $2$ but considering the learning activity instead the social influence. As we can see, the system presents a phase transition from a non-cooperative state to a cooperative one with a low $r_c$ for the three values of $a$ considered. Besides, we observe that $r_c$ increases with $a$. However, we have checked for $a=1$ that the system is not able to overcome the emergence of mutants for the values of $r$ considered. In particular, in this case, we have observed $r_c>10$ and $\bar c$ evolves with strong oscillations. In this way, it is expected an optimal $a \approx 2$ where $r_c$ is minimal. However, it is important to note that the $r_c$ required for cooperation to evolve is larger than the one required considering the social influence. Besides, we observe that the transitions are smoother than the ones shown in the previous section. In particular, the region where $\bar c$ reaches intermediate values correspond to the situation where some realizations finish with a high level of cooperation and other ones near zero cooperation. Besides, for $r>r_c$ we observe that $\bar c$ presents some irregularities where the average fraction of cooperation slightly decreases for some values of $r$. These are produced by the existence of some realization where the cooperative system dies before the $10^4$ generations considered. In this way, we conclude that the learning activity allows to highly cooperative systems overcome the emergence of mutants defectors but with higher $r_c$ and lower life expectancy that the ones obtained through the social influence.

\bigskip

\begin{figure}[!hbt] \centering
\includegraphics[angle=0,scale=0.40]{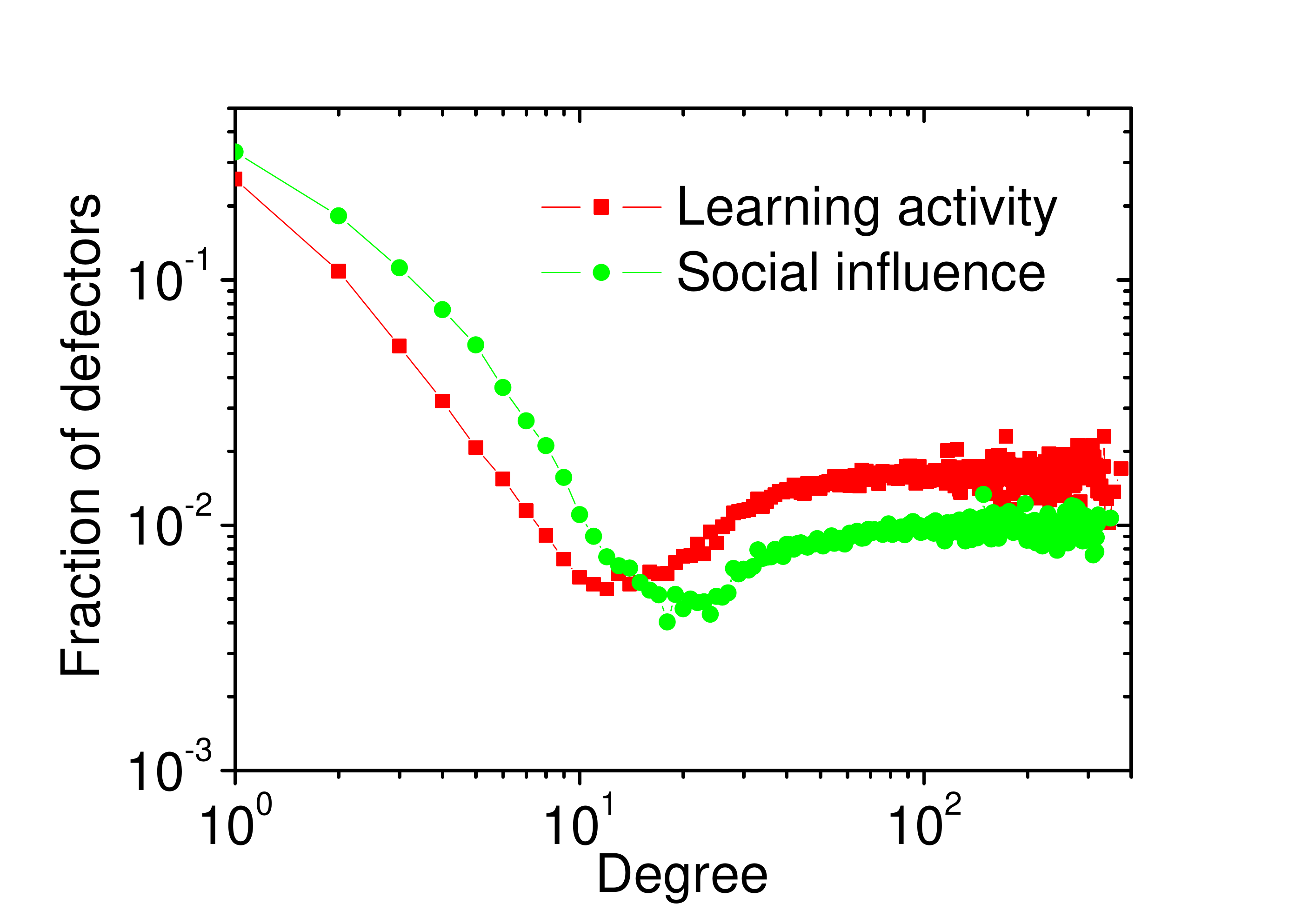}
\caption{Comparison between learning activity and social influence. Fraction of defectors as a function of the degree of individuals. The results have been obtained for $L=4$, $r=4.0$, $P_m=0.01$ and using MA for the network generation. Besides, for the case with social influence we consider $\beta=1$.}
\end{figure}

\bigskip

In order to better understand why the social influence improves the required conditions for cooperation to evolve with respect to the ones obtained with the learning activity, in fig. $8$ we show in log-log scale the fraction of defectors as a functions of the degree of the nodes. Besides, we show the analogous results obtained by considering the social influence to perform a clear comparison. The results have been obtained in the same way that the ones of fig. $6$. In both cases, we observe that defectors preferably inhabit the less linked individuals. However, the fraction of defectors is higher for the case with learning activity than the one with social influence for highly connected individuals. In this way, with the learning activity the system is more prone to have strong defectors and, therefore, the conditions required for cooperation to evolve are larger than the ones required when it is considered the social influence. Nevertheless, although we have shown that social influence have a better performance than the learning activity for the evolution of cooperation, in this section we have also shown the great importance of consider that the abundance of strategies affects the behavior of individuals in highly cooperative systems. However, it would be of great interest perform a thorough theoretical analysis looking for the optimal way to consider the available information in order to ensure the best performance of the focal player.

\bigskip


\bigskip

\begin{thebibliography}{99}

\bibitem{G1} E.O. Wilson, \emph{Sociobiology} (Harvard Univ. Press, Cambridge,
Massachusetts, 1975).

\bibitem{G2} J. Maynard-Smith, E. Szathm\'{a}ry, \emph{The major transitions in
evolution} (Oxford Univ. Press, Freeman, Oxford, 1995).

\bibitem{G3} R.E. Michod, \emph{Darwinian dynamics: Evolutionary transitions in
Fitness and individuality} (Princeton Univ. Press, Princeton, NJ, 1999).

\bibitem{G4} R. Trivers, Q. Rev. Biol. \textbf{46}, 35 (1971).

\bibitem{G5} W.D. Hamilton, J. Theor. Biol. \textbf{7}, 1 (1964).

\bibitem{G6} R. Axelrod, W.D. Hamilton, Science \textbf{211}, 1390 (1981).

\bibitem{G7} M.A. Nowak and K. Sigmund, Nature \textbf{364}, 56 (1993).

\bibitem{G8} R.L. Riolo, M.D. Cohen, and R. Axelrod, Nature \textbf{414}, 441 (2001).

\bibitem{G9} M.A. Nowak and K. Sigmund, Nature \textbf{437}, 1291 (2005).

\bibitem{G10} F.C. Santos, M.D. Santos, and J.M. Pacheco, Nature \textbf{454}, 212 (2008).

\bibitem{G11} M.A. Nowak, Science \textbf{314}, 1560 (2006).

\bibitem{G12} J. Hofbauer and K. Sigmund, \emph{Evolutionary Games and Population Dynamics} (Cambridge University Press, Cambridge, England, 1998).

\bibitem{G13} H. Gintis, \emph{Game Theory Evolving} (Princeton University, Princeton, NJ, 2000).

\bibitem{G14} Kollock, Annu. Rev. Sociol. \textbf{24}, 183 (1998).

\bibitem{R1} G. Szab\'o, G. F\'ath, Phys. Rep. \textbf{446}, 97 (2007).

\bibitem{R2} C. P. Roca, J. A. Cuesta, A. S\'anchez, Phys. Life Rev. \textbf{6}, 208 (2009).

\bibitem{N1} M.A. Nowak and R.M. May, Nature \textbf{359}, 826 (1992).

\bibitem{N2} G. Abramson, M. Kuperman, Phys. Rev. E \textbf{63}, 030901 (2001).

\bibitem{N3} F.C. Santos and J.M. Pacheco, Phys. Rev. Lett. \textbf{95}, 098104 (2005).

\bibitem{N4} F.C. Santos, J.M. Pacheco, and T. Lenaerts, Proc. Natl. Acad. Sci. \textbf{103}, 3490  (2006).

\bibitem{N5} Y.-S Chen, H. Lin and, C.-X Wu, Physica A \textbf{385}, 379 (2006).

\bibitem{N6} S. Assenza, J. G\'{o}mez-Garde\~{n}es, and V. Latora, Phys. Rev. E \textbf{78}, 017101 (2008).

\bibitem{N7} X. Chen, F. Fu, and L. Wang, Physica A \textbf{378}, 512 (2006).

\bibitem{N8} L. Luthi, E. Pestelacci, and M. Tomassini, Physica A \textbf{387}, 955 (2008).

\bibitem{N9} Y.-K. Liu, Z. Li, X.-J. Chen, and L. Wang, Chin. Phys. Lett. \textbf{26}, 048902 (2009).

\bibitem{N13} Z. -X. Wu, J. -Y. Guan, X. -J. Xu, and Y. -H. Wang, Physica A \textbf{379}, 672 (2007).

\bibitem{N10} N. Masuda, Proc R. Soc. B \textbf{274}, 1815 (2007).

\bibitem{N11} A. Szolnoki, M. Perc, and Z. Danku, Physica A \textbf{387}, 2075 (2008).

\bibitem{N12} H. Ohtsuki, C. Hauert, E. Lieberman, and M.A. Nowak, Nature \textbf{441}, 502 (2006).

\bibitem{W1} L. Wardil and J.K.L. da Silva, EPL \textbf{86}, 38001 (2009).

\bibitem{C1} M. Perc and A. Szolnoki, Biosystems \textbf{99}, 109 (2010).

\bibitem{MN}  J. G\'{o}mez-Garde\~{n}es, I. Reinares, A. Arenas, and L. M. Flor\'ia, Sci. Rep. \textbf{2}, 620 (2012).

\bibitem{N20} M. E. J. Newman, SIAM Review \textbf{45}, 167  (2003).

\bibitem{I1} I. Gomez Portillo, Eur. Phys. J. B \textbf{85}, 409 (2012).

\bibitem{I2} I. Gomez Portillo, Phys. Rev. E \textbf{86}, 051108 (2012).

\bibitem{CI} X. Chen, F. Fu, and L. Wang, Physics Letters A \textbf{372}, 1161 (2008).

\bibitem{S1}X. Wang, M Perc, Y. Liu, X. Chen, and Long Wang, Sci. Rep. \textbf{2}, 740 (2012).

\bibitem{S2}A. Szolnoki, Z. Wang, and M. Perc, Sci. Rep. \textbf{2}, 576 (2012).

\bibitem{N18} A.L. Barab\'{a}si, and R. Albert, Science \textbf{286}, 509 (1999).

\bibitem{E1} A. Traulsen, D. Semmann, R. D. Sommerfeld, H.-J. Krambeck, and M. Milinski,  Proc. Natl. Acad. Sci. \textbf{16}, 2962 (2010).

\bibitem{E2} J. Gruji\'c, C. Fosco, L. Araujo, J.A. Cuesta, and A. S\'anchez, PLoS One \textbf{5}, e13749 (2010).

\bibitem{E3} C Gracia-L\'azaro, A. Ferrer, R. Gonzalo, A. Taranc\'on, J.A. Cuesta, A. S\'anchez, and Y. Moreno, Proc. Natl. Acad. Sci. \textbf{109}, 12922 (2012).

\end{thebibliography}
\end{document}